\documentclass[aps, prb, a4paper, notitlepage, reprint, amsfonts, amsmath, amssymb,floatfix]{revtex4-2}

\usepackage{amsmath}
\usepackage[utf8]{inputenc}
\usepackage{graphicx}				
\usepackage{epstopdf}
\usepackage{xcolor}
\usepackage{physics}
\usepackage{float}
\usepackage{hyperref}

\begin{document}

\title{Bilinear-biquadratic spin-1 model in the Haldane and dimerized phases}
\author{Mykhailo V. Rakov}

\affiliation{Department of Physics and Astronomy, Uppsala University, Box 516, S-75120 Uppsala, Sweden}
\affiliation{Institute of Photonics and Quantum Sciences, Heriot-Watt University, Edinburgh EH14 4AS, United Kingdom}

\author{Michael Weyrauch}
\affiliation{Physikalisch-Technische Bundesanstalt, Bundesallee 100, D-38116 Braunschweig, Germany}

\date{\today}

\begin{abstract}
We study the low-lying spectrum of the bilinear-biquadratic Heisenberg model in the dimerized and Haldane phases using a tensor renormalization method. At the critical point $\theta=-\pi/4$ the finite-size spectrum predicted by the Wess-Zumino-Witten (WZW) model can only partly be confirmed. We find a singlet-singlet gap which does not fit into the WZW systematics.  The results obtained are compared to Bethe Ansatz, exact diagonalization, and DMRG calculations for specific parameters.
\end{abstract}
\maketitle

\section{Introduction}

The isotropic bilinear-biquadratic (BLBQ) Heisenberg model plays a fundamental role in the theory of magnetism, and, more generally,
 for the understanding of strongly interacting many body systems. For spin-1 systems in one spatial dimension the model is given by the Hamiltonian
\begin{equation}\label{eq-biqui}
{\cal H}=\sum_{i=1}^N \cos \theta (\vec{S}_i \otimes  \vec{S}_{i+1}) +
\sin \theta (\vec{S}_i \otimes\vec{ S}_{i+1})^2.
\end{equation}
The $S^i_\lambda$ are spin-1 matrix representations of SU(2) and $N$ denotes the system size.
The model depends on the parameter $\theta$, which governs the ratio between the bilinear and biquadratic terms.
In the present contribution we concentrate on periodic one-dimensional (1D) systems with nearest neighbour interactions only.

\begin{figure}
\unitlength1cm
\begin{picture}(18,6.5)(0,0)
 \put(1,0)  {\includegraphics[width=5.5cm]{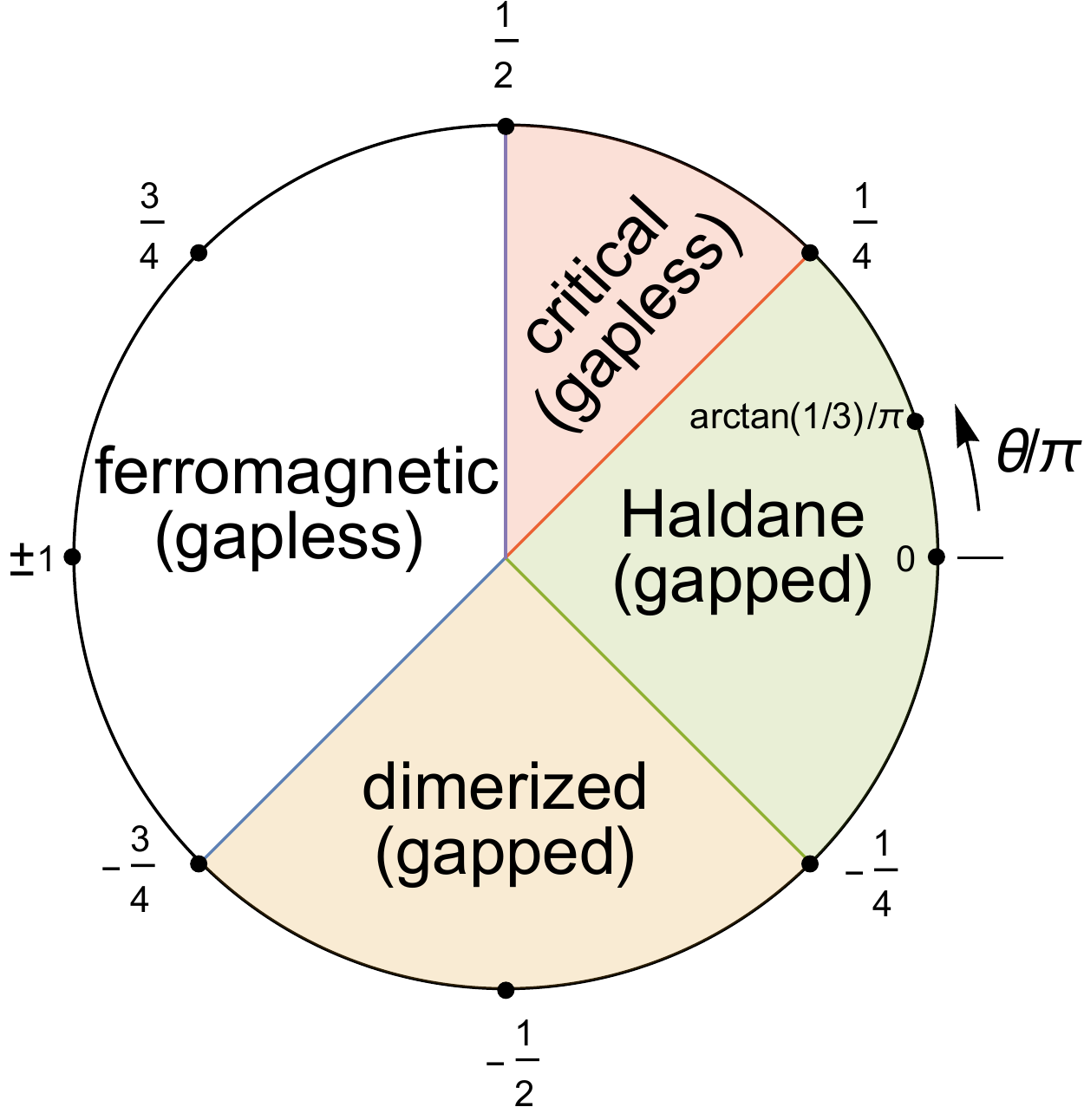}}
\end{picture}
\caption{\footnotesize Phase diagram of the spin-1 bilinear-biquadratic (BLBQ) Heisenberg model as a function of $\theta$.
There are four quantum phases: the ferromagnetic phase, the critical (trimerized) phase, the Haldane phase, and the dimerized phase.
\label{fig:phaseDiagram}}
\end{figure}

The 1D BLBQ model shows a  rich phase structure with various disordered phases. The 1D  phase diagram
differs from the one expected for higher dimensions~\cite{PhysRevB.51.3620} and is shown schematically in Fig.~\ref{fig:phaseDiagram} as a function
of the parameter $\theta$. One finds three exotic phases: the massless trimerized phase, the Haldane phase, and
the dimerized phase. In fact, in one dimension, due to Coleman's theorem~\cite{Coleman73}, the only ordered phase is ferromagnetic since its order parameter $S_z$ is conserved.

This phase structure should be reflected in the excitation spectrum of low lying states,
with gaps closing at the critical points $\theta=\pm\pi/4$, which separate the disordered phases from each other.
Here we will study the spectrum on both sides of the critical point at $\theta=-\pi/4$, specifically from $\theta=-\pi/2$
to $\theta=0$. The spectrum in the whole region should be gapped except at the critical point. The gaps are expected to be small and are not easily determined precisely.

Previously, the ground state of the model was studied using a variational method in Ref.~\cite{PhysRevB.101.235145}. For the excitation spectrum there are
rather old exact diagonalization results~\cite{PhysRevB.51.3620}, which encompass the Haldane and dimerized phases. Furthermore, there is a
systematic study of the excitation spectrum of the BLBQ model in the Haldane phase in Ref.~\cite{PhysRevB.88.075133} using tensor renormalization methods with relatively small tensor sizes
as well as various studies at
isolated points in the Haldane phase ($\theta=0$~\cite{WHI93a,PhysRevB.85.100408} and AKLT point $\theta=\arctan{\frac{1}{3}}$~\cite{AFL1987}).
 Moreover, there are  Bethe Ansatz
results for $\theta=-\pi/2$~\cite{BAR89,SOR90} and for the critical point $\theta=-\pi/4$~\cite{TAKHTAJAN1982479,BABUJIAN1982479,Alcaraz_1988, AlcarazMartins1988}.

We use the higher-order tensor renormalization group (HOTRG) method~\cite{XIE2012} to determine the low-lying spectrum.
Our variant of this method  implements U(1) symmetry of the tensors
explicitly.
Tensor renormalization is able to obtain spectra for relatively large systems, and
in a recent paper~\cite{PhysRevB.100.134434} we studied the XXZ
chain in a longitudinal homogeneous field and showed by comparison to Bethe ansatz results that the method
 accurately determines the spectrum and the phase diagram.

The low lying spectrum is calculated from the transfer matrix, which is
obtained from the coarse grained tensors. The renormalized tensor at each coarse graining step corresponds
to a certain system size. Therefore, one obtains the complete finite size dependence of the spectrum required for the
determination of the scaling dimensions in a single run of the tensor renormalization procedure.

It is our goal to obtain the low lying spectrum as precisely as possible from our numerics and then compare it to various other
analytical and numerical results.

\section{The critical point $\theta=-\pi/4$}\label{sec-finite-size}

We start with an investigation of the gapless spectrum at the critical point $\theta=-\pi/4$.
At the critical point the spin-1 BLBQ model (\ref{eq-biqui}) may be mapped to a Wess-Zumino-Witten (WZW) model with symmetry group SU(2) and topological index $k=2$~\cite{AFFLECK1986409, PhysRevLett.56.2763}.
This model is characterized by a central charge
$c=3/2$, and the  scaling dimensions of the primary fields $x_0=0$, $x_{1/2}=3/8$, and $x_1=1$. The index corresponds to the spin of the primary field
with $x_j=2j(j+1)/(2+k)$ and $j\leq k/2$~\cite{KNIZHNIK198483}.

Since the WZW model is conformally symmetric, the ground state energy  should scale with system size $N$
as~\cite{PhysRevLett.56.742, PhysRevLett.56.746}
\begin{equation}\label{gs}
\frac{E_0}{N} = e_0 - \frac{\pi c v_s}{6 N^2}  + O[N^{-2} (\log N)^{-3} ]
\end{equation}
with $e_0$ the ground state energy per site of the infinite system and the spin wave velocity $v_s$. We have indicated
logarithmic corrections due to the marginal operator with scaling dimension $x_j>1$~\cite{Cardy_1986}. These corrections are included
in the fits to our numerical data.

The excitation gaps $\Delta_{j n n^\prime}$ can be parameterized by the scaling dimensions $x_j$
\begin{equation}\label{delta}
\Delta_{j n n^\prime}=\frac{2\pi v_s}{N} (x_j+n+n^\prime)  + O[N^{-1}(\log N)^{-1})]
\end{equation}
with $n,n^\prime= 0,1,2, \ldots$.

Moreover, at the critical point the BLBQ spin model may be solved  using the
Bethe Ansatz~\cite{TAKHTAJAN1982479, BABUJIAN1982479, Alcaraz_1988, AlcarazMartins1988}.  This enables the determination of the non-universal quantities $e_0=-2\sqrt{2}$ and  $v_s=\pi\sqrt{2}$.
Combining these results with the central charge and scaling dimensions of the WZW model completely determines the finite size spectrum up to linear order in $1/N$.

\begin{figure}
\unitlength1cm
\begin{picture}(9,10.)(0,0)
 \put(.2,5.2)  {\includegraphics[width=8.cm]{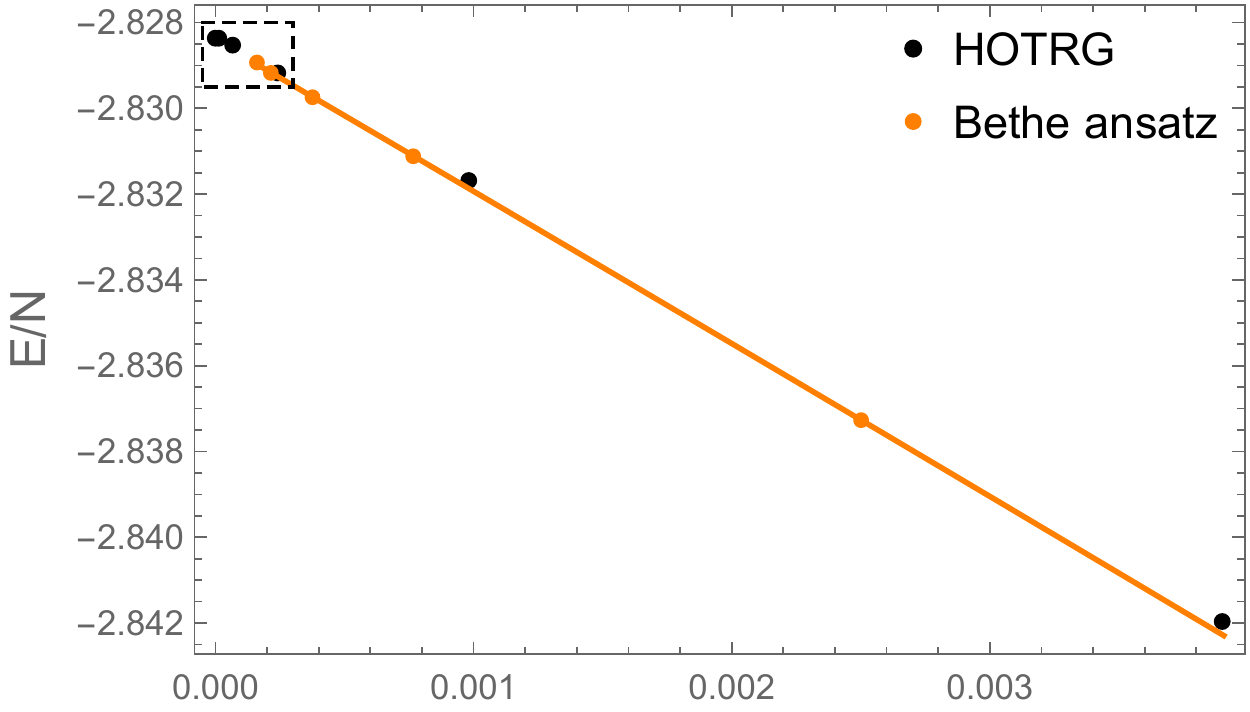}}
 \put(0,0)      {\includegraphics[width=8.3cm]{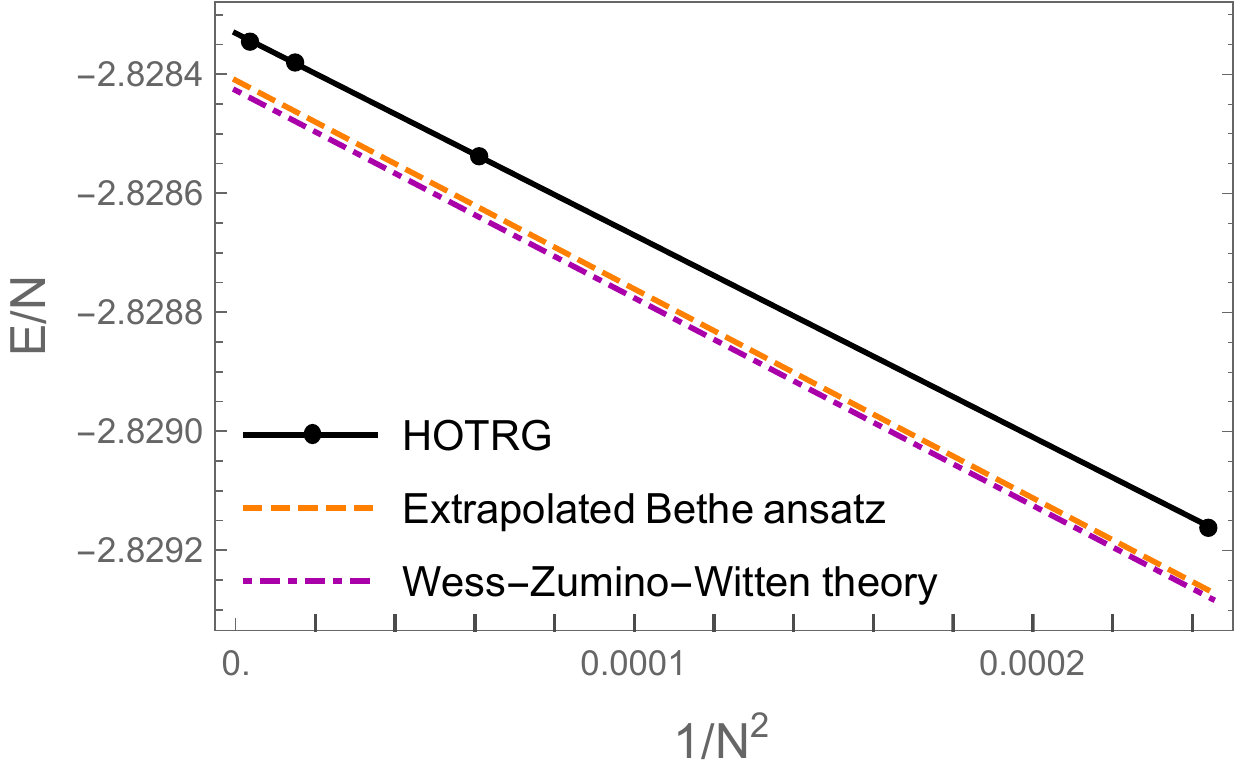}}
\end{picture}
\caption{\footnotesize Ground state energy per site at the critical point $\theta=-\pi/4$ as a function of $1/N^2$.
Upper panel: HOTRG results (black dots) are compared to Bethe Ansatz data (orange dots)~\cite{Alcaraz_1988}. The fit to the Bethe Ansatz data (orange curve) includes logarithmic corrections
according to Eq.~(\ref{gs}): ${E_0}/{N}=-2.82841 - 3.49292 /N^2 -1.29507/[N^2 (\log N)^3]$.
Lower panel: the HOTRG data in the dashed rectangle indicated in the upper panel are plotted again at larger scale for system sizes $N\geq64$ (black dots).
The black line shows a fit to all HOTRG data including logarithmic corrections according to Eq.~(\ref{gs}): ${E_0}/{N}=-2.82833 - 3.41099/N^2 - 1.69186/[N^2 (\log N)^3]$.
The dashed orange line indicates an extrapolation of the Bethe ansatz data using the fit given above.
\label{fig:energyComparison}}
\end{figure}

HOTRG numerical data  are compared to finite size Bethe Ansatz results for small systems ($N\leq 84$) ~\cite{Alcaraz_1988, AlcarazMartins1988} in Fig.~\ref{fig:energyComparison}.
HOTRG data for larger systems with $N\geq 64$ are shown again  in the lower panel of this figure at larger scale.
From the fit to the HOTRG data  we obtain $e_0=-2.82833$ and
$c v_s=6.48$. Both values agree rather well with the expectations $e_0=-2\sqrt{2}\approx -2.82843$ and $c v_s=3\pi/\sqrt{2}\approx 6.664$. The extrapolated Bethe Ansatz data
depicted in Fig.~\ref{fig:energyComparison} provide $e_0=-2.82841$ and
$c v_s=6.670$. We attribute the small systematic discrepancies to the finite imaginary time step $\tau=0.004$ used for the HOTRG calculations as well as other (numerical) approximations
which are discussed in some detail in Ref.~\cite{PhysRevB.102.104422}.

\begin{figure}
\unitlength1cm
\begin{picture}(18,12.)(0,0)
 \put(.3,5.6)  {\includegraphics[width=8.cm]{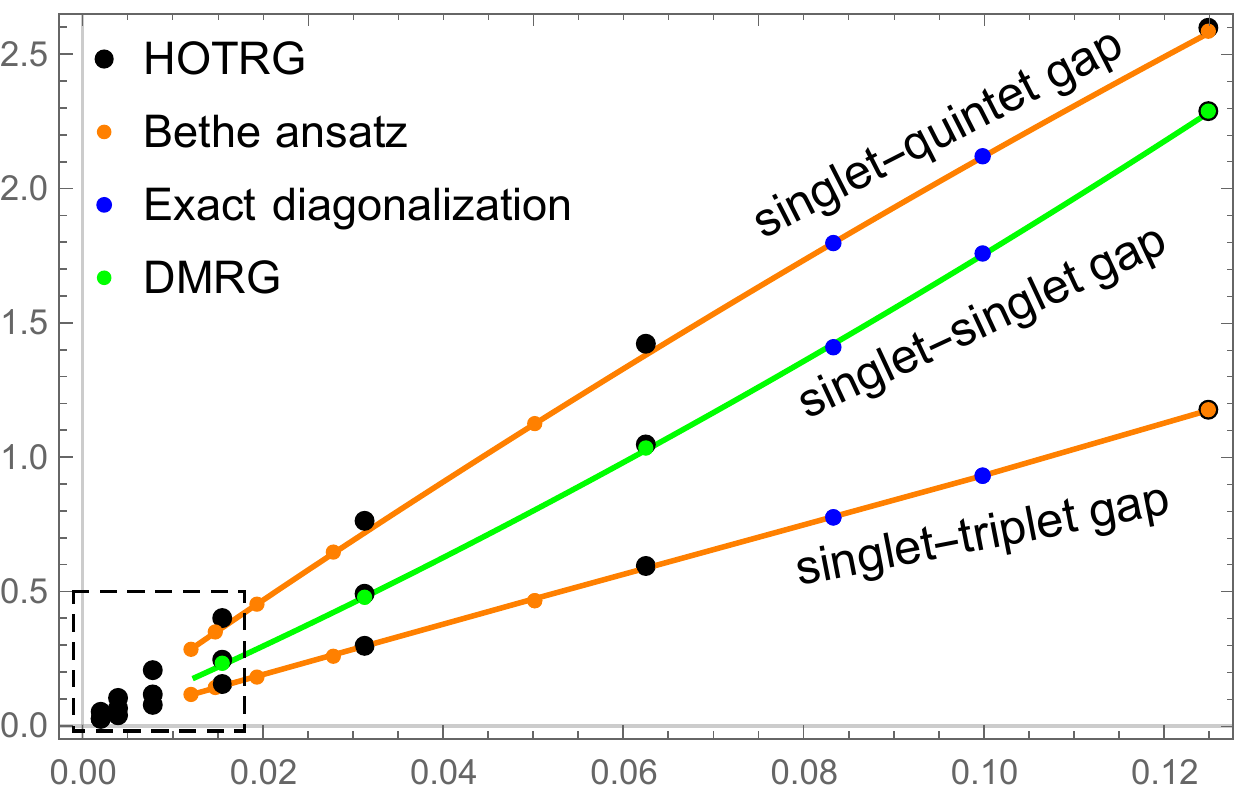}}
 \put(.3,0)    {\includegraphics[width=8.cm]{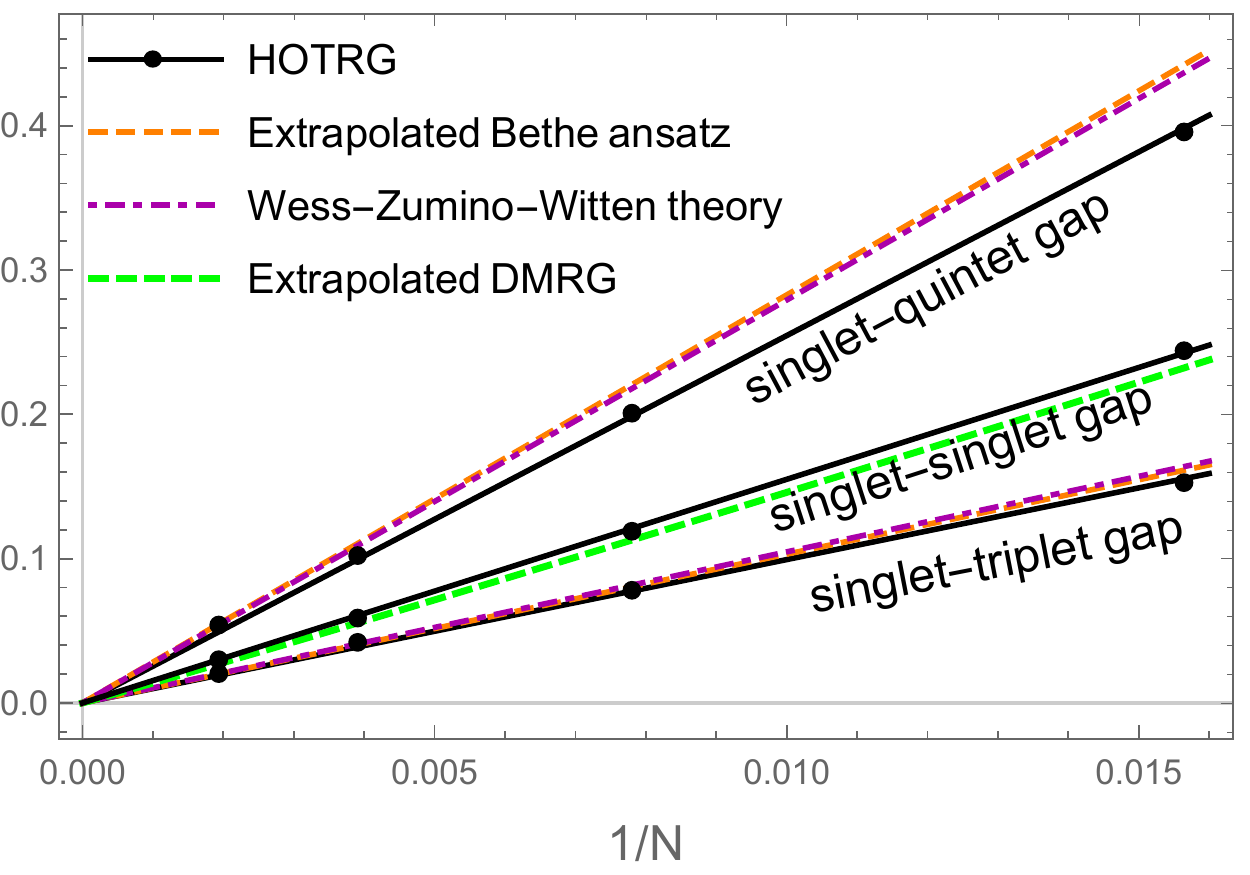}}
\end{picture}
\caption{\footnotesize Spectral gaps in various SU(2) symmetry sectors at the critical point $\theta=-\pi/4$ as functions of $1/N$. Upper panel: HOTRG data compared to Bethe Ansatz data~\cite{Alcaraz_1988, AlcarazMartins1988}, exact diagonalization~\cite{PhysRevB.51.3620} and DMRG results. The fits to Bethe ansatz data~\cite{Alcaraz_1988} (orange curves) and DMRG/exact diagonalization data (green curves) include logarithmic corrections.  Lower panel: The HOTRG data for system sizes $N\geq 64$ compared to extrapolations of the fits as well as WZW model predictions. The parameters used for the extrapolations are
for the triplet gap: $9.59527/N - 0.406537/(N \log N)$;
for the singlet gap: $10.8557/N + 15.4145/(N \log N)$;
for the quintet gap: $26.4171/N - 12.0204/(N \log N)$.
\label{fig:gapsComparison}}
\end{figure}

HOTRG data for the gaps at the critical point are shown in Fig.~\ref{fig:gapsComparison}. They are compared to Bethe Ansatz results ($N<84$)~\cite{Alcaraz_1988, AlcarazMartins1988}, exact diagonalization data~\cite{PhysRevB.51.3620} for small systems ($N=10$ and $N=12$) as well as DMRG data ($N=16$,$N=32$ and $N=64$) we calculated with a code~\cite{RakovWeyrauch2017} explicitly implementing SU(2) symmetry.
Two different fits  to the HOTRG data based on Eq.~\ref{delta} are presented in Tab.~\ref{gapfit} together with results for the WZW model.
The singlet-triplet and singlet-quintet coefficients reasonably match the expectations of the WZW model, if we assume the scaling dimensions given in the table.
Again there is a small systematic deviation between Bethe Ansatz and HOTRG data, which we attribute to the numerical approximations alluded to above for the ground state energy.

Surprisingly, the calculated singlet-singlet gap does {\it not} fit into the expected WZW model systematics. While we find the expected linear scaling with $N^{-1}$, we cannot
assign a WZW model scaling dimension to this gap. The numerical data which include also data points from DMRG calculations clearly predict the singlet-singlet gap
above the singlet-triplet gap and below the singlet-quintet gap with sufficient precision in order to reach this conclusion. To our knowledge there are no results from Bethe Ansatz for this gap.

\begin{table}[t]
\caption{\label{gapfit} Coefficients $C$ and $C^\prime$ of the finite-size expansion for the gaps $\Delta=\frac{C}{N} + \frac{C^\prime}{N \log N}$. For the WZW model the coefficients are calculated
from Eq.~(\ref{delta}) using $c=3/2$ and $v_s=\pi\sqrt{2}$ with the scaling dimensions $x$ given in the table. (t=triplet, s=singlet, q=quintet). HOTRG Fit 1 uses {($64\leq N \leq 512,~C^\prime=0$)} and HOTRG Fit 2 uses {$(8\leq N \leq 512)$ }}.
\begin{ruledtabular}
\begin{tabular}{|l| r r|r| r r|}
    &  \multicolumn{2}{c|} {WZW} & \multicolumn{1}{c|} {HOTRG} & \multicolumn{2}{c|} {HOTRG }\\
    &                  &      &  \multicolumn{1}{c|}   {Fit 1} & \multicolumn{2}{c|}  {Fit 2}      \\

    & $x$            & {$C$} &  {$C$} & $C$ & $C^\prime$ \\
  \hline
  t & $\frac{3}{8}$  &10.5   & 9.9    &  10.0 & -1.2     \\
  s &                &       & 15.5   &  12.2 & 12.5     \\
  q & $1$            & 27.9  & 25.5   &  29.1 & -17.2    \\

\end{tabular}
\end{ruledtabular}
\end{table}

\section{Ground state energy and spectral gaps}\label{sec-XXZ}

\begin{figure}
\unitlength1cm
\begin{picture}(18,6.5)(0,0)
 \put(0,0)  {\includegraphics[width=8.5cm]{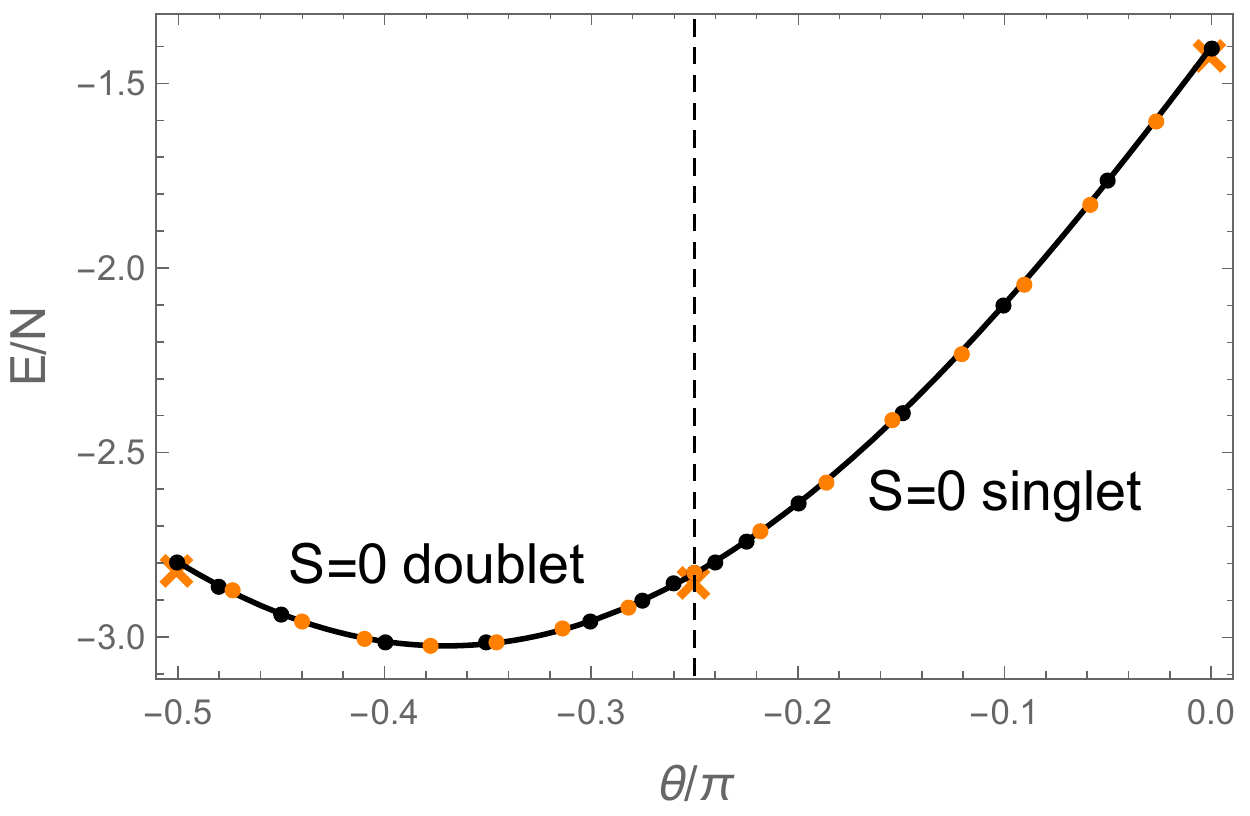}}
\end{picture}
\caption{\footnotesize Ground state energy per site as a function of $\theta$. The orange points are results
obtained in Ref.~\cite{PhysRevB.101.235145} using DMRG. The orange crosses at $\theta=-\pi/2$ and $\theta=-\pi/4$ are the Bethe ansatz result from Ref.~\cite{SOR90} and~\cite{TAKHTAJAN1982479}, respectively, while the value at $\theta=0$ is the DMRG result from
from Ref.~\cite{WHI93a}.
\label{fig:gs}}
\end{figure}

Now we present results of the low lying spectrum in the parameter region
from $\theta=-\pi/2$ to $\theta=0$. This region includes the critical point
discussed in section~\ref{sec-finite-size}, and covers parts of the dimerized and Haldane phases.
As SU(2) symmetry is not broken, all states come in corresponding multiplets. For reference, in Fig.~\ref{fig:gs} we show the energy of the $S=0$ ground state as a function of $\theta$ obtained from our HOTRG calculations and compare it to results from various other methods.

In the dimerized region the next singlet state (which was discussed for $\theta=-\pi/4$ in the previous section) is degenerate
with the ground state singlet. These two $S=0$ states form the dimer. Numerically the energy splitting is
$10^{-2}$ or smaller. 
In the Haldane phase the two singlets split up, and the excited singlet is found in the calculated spectrum above several triplet states.
Unfortunately, it is rather difficult to detect this singlet in the `sea' of triplets which exists between
the lowest triplet and the quintet for $\theta>-\pi/4$ as discussed in more detail below.

\begin{figure}
\unitlength1cm
\begin{picture}(18,10.)(0,0)
 \put(0,4.8)  {\includegraphics[width=8.5cm]{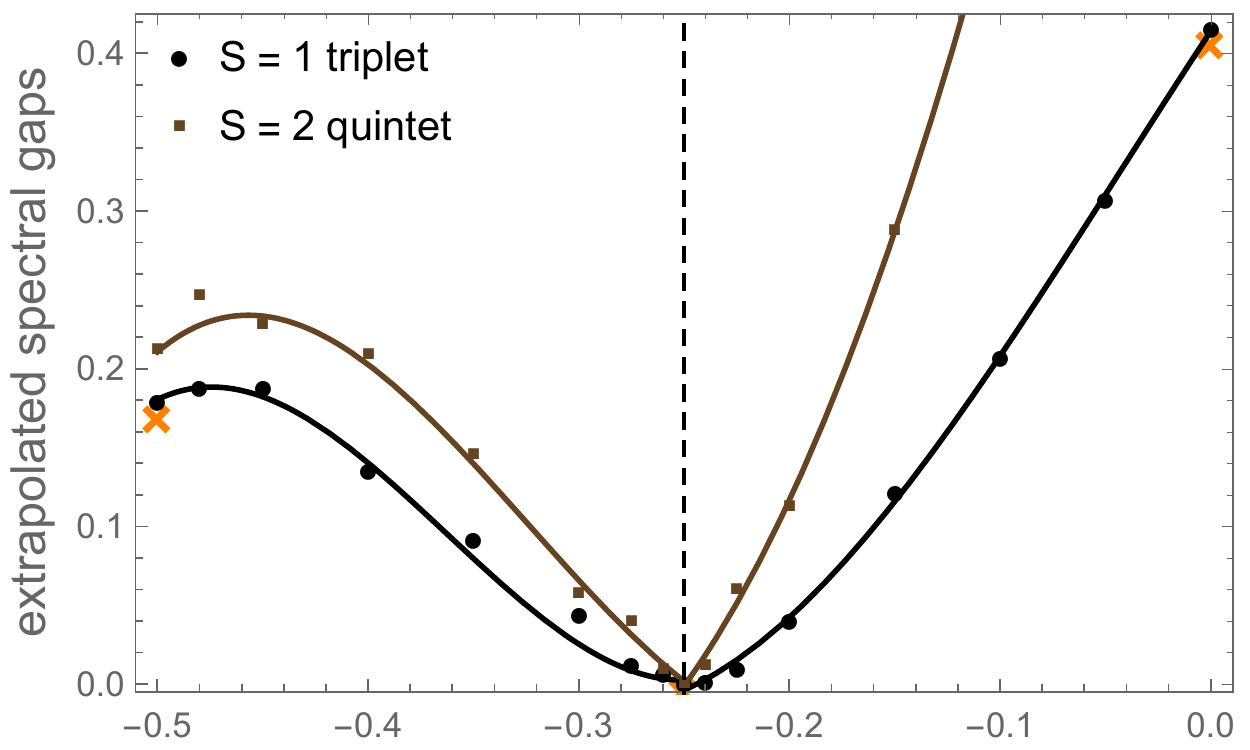}}
 \put(3.8,0)  {\includegraphics[width=4.55cm]{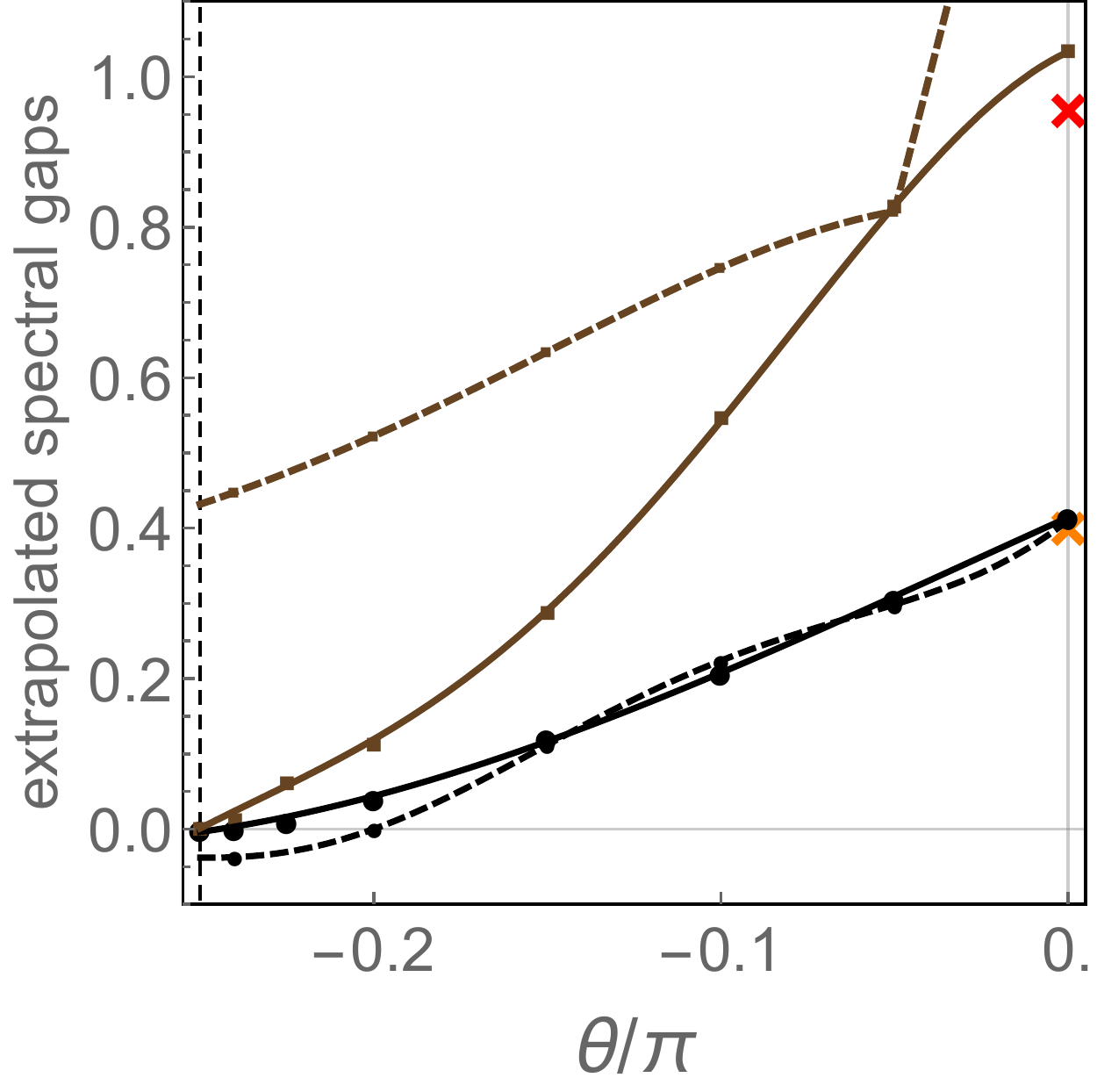}}
\end{picture}
\caption{\footnotesize  Lowest singlet-triplet and singlet-quintet gaps as functions of $\theta$.
 Upper panel: $S=1$ triplet (large dots), $S=2$ quintet (small dots). The orange crosses denote the Bethe ansatz result at $\theta=-\pi/2$ from Ref.~\cite{SOR90} and DMRG results at $\theta=0$ from Ref.~\cite{WHI93a,PhysRevB.85.100408}, respectively.
 The solid curves are guides to the eye.
 Lower panel: details in the Haldane phase, comparison of our HOTRG results (solid curves) to results from Ref.~\cite{PhysRevB.88.075133} (dashed curves) obtained with MPS size $m=24$. The crosses indicate the results from~Ref~\cite{PhysRevB.85.100408} obtained with MPS size $m=30$. 
 The result for the singlet-quintet gap obtained in Ref.~\cite{PhysRevB.88.075133} at $\theta=0$ is about 1.72.
\label{fig:gap}}
\end{figure}

Numerical data for the lowest SU(2) triplet and quintet gaps are shown in Fig.~\ref{fig:gap}. 
For comparison we include data  obtained with other methods as annotated in the figure caption. 
It is important to note that we only show the {\it lowest} triplet and quintet. Many more closely spaced triplet states  are found in the spectral region between
the lowest triplet and quintet, and similarly many more quintet states  are found  above the lowest one. All these states correspond to different quasi-momenta.
At the critical point the lowest triplet and quintet show cusps. Whether these cusps follow  $\Delta \sim |\theta-\theta_c|$ as was suggested by
Affleck~\cite{AFFLECK1986409} cannot be determined precisely from our data.

 A systematic study of the low-lying spectrum of infinite-size BLBQ systems in the Haldane phase was presented in Ref.~\cite{PhysRevB.88.075133} using matrix product states (MPS) of size $m=24$. Corresponding data  are shown in the lower panel of~Fig.~\ref{fig:gap} and compared to our HOTRG results. 
 For the triplet gap both sets of data compare rather well.
Only in the vicinity of the critical point are the results of~\cite{PhysRevB.88.075133} numerically imprecise as the gap becomes negative. The high precision result~\cite{PhysRevB.85.100408} for the triplet gap at the Heisenberg point $\theta=0$ (obtained with an MPS size $m=30$) may serve as a benchmark for the precision of the calculations.

The situation is more complicated for the quintet gap: here our result at $\theta=0$ is about $7\%$ larger than the data point given in Ref.~\cite{PhysRevB.85.100408}, but the data provided in Ref.~\cite{PhysRevB.88.075133} differ quite drastically in the whole Haldane phase from our results.  In particular, we  cannot confirm the jump of the quintet observed in Ref.~\cite{PhysRevB.88.075133} at the critical point $\theta=-\pi/4$. We find a cusp instead as for the triplet.

\begin{figure}
\unitlength1cm
\begin{picture}(18,6.5)(0,0)
 \put(0,0)  {\includegraphics[width=8.5cm]{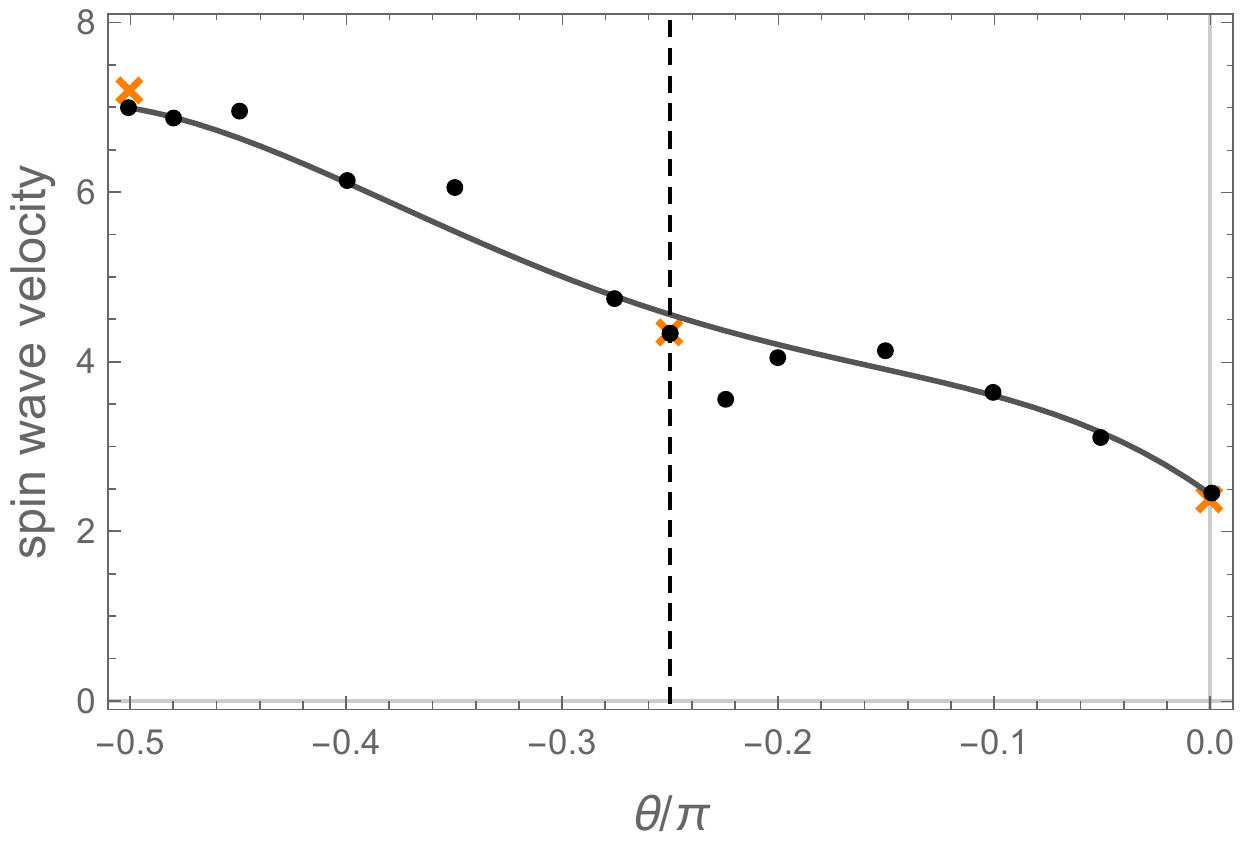}}
\end{picture}
\caption{\footnotesize Spin wave velocity as a function of $\theta$, obtained as the product of the spectral gap $\Delta$ and the correlation length $\xi$. At $\theta=-\pi/4$ the velocity is obtained from finite-size scaling of the energy per site in good agreement with Bethe ansatz prediction $v_s=\pi\sqrt{2}$ (red cross). The grey curve is guide to the eye.
\label{fig:velocity}}
\end{figure}

From the triplet gap $\Delta$ and the correlation length $\xi$ determined in Ref.~\cite{PhysRevB.101.235145} we calculate the spin wave velocity
$v_s= \xi \Delta$ which is shown in Fig.~\ref{fig:velocity}. The grey curve is a guide to the eye. At the critical point $\theta=-\pi/4$ (where $\Delta=0$ and $\xi = \infty$) the velocity can be obtained from finite-size scaling of the energy per site, which is in line with Bethe Ansatz prediction $v_s=\pi\sqrt{2}$ as indicated in section~\ref{sec-finite-size}.

The spin wave velocity depends weakly on $\theta$. We are able to reproduce with good precision the velocity at  $\theta=-\pi/2$ and $\theta=0$, where results of other calculations are available
as annotated in the figure caption.
The precision of the presented data becomes questionable in the vicinity of the critical point $\theta=-\pi/4$. Firstly, precise extrapolation of the gap to the thermodynamic limit becomes very difficult in this area. Secondly, the velocity data also depend on the precision of the results for the correlation length obtained in Ref.~\cite{PhysRevB.101.235145}.

\section{Conclusion}

We studied the low-lying spectrum of the bilinear-biquadratic Heisenberg model in the dimerized and Haldane phases using a tensor renormalization method. The spectral gaps were numerically calculated with high precision in a large parameter region, and we found good agreement with Bethe Ansatz and DMRG results as well as exact diagonalization wherever available. The spin wave velocity was also calculated.

The finite-size spectrum predicted by the Wess-Zumino-Witten (WZW) model at the critical point $\theta=-\pi/4$ was confirmed with regard to triplets and quintets. However, we find a singlet-singlet gap which does not fit into the WZW systematics. The latter result is rather surprising and requires further investigation.

\acknowledgments
M.V.R. acknowledges the support of the Olle Engkvist Byggm\"astare Foundation under Decision No. 198-0389. The computations were partly performed on resources provided by SNIC through Uppsala Multidisciplinary Center for Advanced Computational Science (UPPMAX) under Projects SNIC 2020/3-20 and SNIC 2021/22-172, and National Supercomputer Center (NSC) under Project SNIC 2021/1-23.
Computing time was as well provided by Physikalisch-Technische Bundesanstalt (PTB).

\end{document}